\documentclass[12pt,pdftex]{article}
\usepackage[dvips]{graphicx}
\usepackage{amsmath,amssymb,amsthm}
\usepackage{mathbbol,bbm}
\usepackage[final]{showkeys}
\usepackage{bm}
\usepackage{calc}

\begin{document}

\begin{center}
{\LARGE Quantum memory as a perpetuum mobile? Stability v.s. reversibility of information processing.} \\[18pt]
R.~Alicki\\[6pt]
Institute of Theoretical Physics and Astrophysics \\
University of Gda\'nsk, Poland \\[6pt]
and Weston Visiting Professor\\[6pt]
Weizmann Institute of Science, Rehovot, Israel\\[6pt]
\end{center}

\begin{abstract}
It is argued using a Gedankenexperiment that a scalable quantum memory could be used as a perpetuum mobile of the second kind and hence cannot be realized in Nature.
The reasoning is based on the assumption that the Landauer's principle for measurements is a consequence of the second law of thermodynamics and not an independent postulate. This implies a modification of the Landauer's principle when applied for discrimination of equilibrium (metastable) states. While identification of the metastable state can be done at the infinitesimally low cost, a change of such a state involves dissipation of energy proportional to its stability factor.  
\end{abstract}

\section{Introduction}
This note concerns the fundamental question in quantum information:\emph{ Is fault-tolerant Quantum Information Processing (QIP) feasible?}\\
The extensively studied and well-developed theory of fault-tolerant quantum computation \cite{ZurekFT,DoritFT,Gottesman2000-FT-local,Preskill1998-rel}
provides a positive answer, however its phenomenological assumptions are doubtful and often criticized \cite{AHHH2001,RAFNL,AlickiLZ2006-FT,Dya}. On the other hand first principle models are very difficult to analyze and do not give a complete and general answer yet. Therefore, one can ask an easier question:\emph{ Is a scalable quantum memory feasible?}\\
In the recent years several models of quantum memory based on  self-correcting spin systems
have been proposed \cite{K,DennisKLP2002-4DKitaev,Bacon} and few of them rigorously analyzed \cite{AlickiFH2007-Kitaev,AlickiFH2008-Kitaev,AlickiH3-Kitaev,Ter,Hor}. Despite certain stability properties, proved or expected  for some of those models, there is no proof that any of the proposals satisfies all the needed conditions for scalable quantum memory.\\
A natural question arises: \emph{Can phenomenological thermodynamics provide restrictions or even no-go theorems for Quantum Memory, or generally for QIP ?}\\
An attempt of \cite{AlickiH2006-drive} based on the KMS theory was not convincing for many experts because it was based on the mathematical structures related to thermodynamical limit in terms of quasi-local algebras. Here another, more heuristic, approach is presented  involving the modified Landauer's Principle for quantum measurements and a \emph{Gedankenexperiment} on a system implementing quantum memory. 
\par
The main problem with thermodynamical arguments is that the laws of thermodynamics are usually formulated in a natural language and have a common sense character.
To apply them to some subtle problems one needs more rigorous formulations, than those found in the most textbooks. This is particularly important in the quantum theory, which often seems to be far from a "common sense".\\
Another problem is the question of applicability of thermodynamics to QIP. There exist two points of view:

1) Physical systems used for QIP are different from those considered in thermodynamics and therefore thermodynamical restrictions do not apply.

2) Thermodynamics applies.

The author of the present note shares the second opinion following the famous statements : \\
\emph{But if your theory is found to be against the second law of thermodynamics I can give you no hope; there is nothing for it but to collapse in deepest humiliation.} - Sir Arthur Stanley Eddington.\\
\emph{ (Thermodynamics)...is the only physical theory of universal content which I am convinced that within the framework of applicability of its basic concepts, it will never be overthrown.} - Albert Einstein.

\section{Applicability of thermodynamics to information processing}

The laws of thermodynamics possess a phenomenological and common sense character. In particular the limits of their applicability
lie beyond the scope of phenomenological thermodynamics and need serious considerations based on first principle microscopic models. For example one can take a possible formulation of the \emph{Zero-th Law}:\\
\emph{Any system coupled to a thermal bath relaxes to the thermodynamical equilibrium state at the bath's temperature},\\
and the \emph{Second Law}:\\
\emph{It is impossible to obtain a process such that the unique effect is the subtraction of a positive heat from a reservoir and the production of a positive work.}\\
In both cases one can ask the questions: How long does thermal relaxation or subtraction of heat can take? Are there any relations between the size of the systems and the time scale of those processes? Similarly, what should be a scale of produced work and how should it depend on the size of the system? \\
\par
As a simple example consider a ferromagnet consisting of $N$ microscopic constituents ("spins") below the critical temperature. In principle, any polarized macroscopic state ultimately relax to the unique "equilibrium state" for which the direction of magnetization ${\vec M}$ is completely unpredictable. The states with fixed direction of ${\vec M}$ are metastable  with relaxation times  increasing \emph{exponentially} with $N$. For large $N$ such metastable states possess all expected features of equilibrium states and moreover the rigorous approach involving thermodynamical limit treats them as true equilibrium states corresponding to pure thermodynamical phases. To find an additional relation between admissible time scales of thermodynamic processes one can be guided by the analogous problems in computer science. In the theory of complexity the problem can be solved efficiently if the time needed for the solution grows polynomially with the input size. This suggests the following reformulation of the \emph{Zero-th Law}:\\
\emph{Any $N$-particle system coupled to a thermal bath relaxes to the (possibly nonunique) thermodynamical equilibrium state at the bath's temperature with relaxation time growing at most polynomially in $N$},\\
and the \emph{Second Law}:\\
\emph{It is impossible to obtain an effective process such that the unique effect is the subtraction of a positive heat from a reservoir and the production of a positive work of the order of at least $kT$. The effective process means a process which takes at most polynomial time in the number of particles $N$.}\\
The study of applicability of thermodynamics to classical and quantum systems reached a mature status quite recently with the development of \emph{fluctuation theorems} \cite{Fluctuation}. The rough formulation of the fluctuation theorem is the following:\\
\emph{For a system consisting of $N$ particles the probability of observing during time $t$ an entropy production opposite to that dictated by the second law of thermodynamics decreases exponentially with $Nt$.}\\
Such a formulation has an immediate consequence for classical and quantum computing. Namely, one cannot hope that the efficiency of any computing scheme can follow from the hypothetical deviations from phenomenological thermodynamics. 

\section{Measurements and Landauer's Principle}

Landauer's Principle, first argued in 1961 by Rolf Landauer of IBM \cite{Landauer}, holds that \emph{any logically irreversible manipulation of information, such as the erasure of a bit or the merging of two computation paths, must be accompanied by a corresponding entropy increase in non-information bearing degrees of freedom of the information processing apparatus or its environment} \cite{Bennett}.

Specifically, each bit of lost information will lead to the release of an amount $kT\ln 2 $ of heat, where $k$ is the Boltzmann constant and $T$ is the absolute temperature of the circuit. On the other hand, if no information is erased, computation may in principle be achieved which is thermodynamically reversible, and require no release of heat. This has led to a considerable interest in the study of reversible computing.\\

The Landauer's Principle has a direct relation to thermodynamics of  quantum measurement processes. Here, a \emph{quantum measurement} is a projective von Neumann measurement of an observable $ A = \sum_j a_j P_j$ such that for an initial state $\rho$ of the system the final state after measurement is given by:
\begin{equation}
\rho_k = \frac{P_k \rho P_k}{\mathrm{Tr}(P_k \rho P_k)}
\label{meas}
\end{equation}
when the outcome $a_k$ is recorded.
In order to use the measurement's outcome for the system's control one has to assume that after a measurement the system remains in the corresponding
state for the time at least of the order $\mathcal{O}(1)$.\\
Consider a 2-level quantum system with a trivial initial Hamiltonian $H(t_0) = 0$ coupled to a heat bath at the temperature $T$. One can design a cyclic procedure of extracting work from a bath consisting of the following steps \cite{ALH32004}:\\
i) Measurement on the system at the initial state $\rho(t_0)= I/2$ in the basis of $\sigma^z$ which yields the outcome 
$s= \pm 1$ with a collapsed state $\rho (t_1)= |s\rangle\langle s|$.\\
ii) Fast (in comparison to the thermal relaxation time) switching on an external field producing the Hamiltonian  $H_s(t_1) = (s E/2)(sI-\sigma^z )$ which increases the energy of the state $|-s\rangle$ by $E >> kT$ and  does not change the energy of
the state $|s\rangle$.\\
iii) Slow (again in comparison to the thermal relaxation time) switching off the external fields such that $H(t_2) = 0$. Reseting of the measuring device.\\
\par
One can  compute the balance of work $W(t)$, heat $Q(t)$ and  internal energy $E(t) $ during the full cycle $t_0\to t_1\to t_2$
using the basic definitions discussed in \cite{AL1979,ALH32004} and recalled in the Appendix I
\begin{equation}
E=\mathrm{Tr }(\rho H),\quad dW = \mathrm{Tr }(\rho\, dH), \quad dQ =\mathrm{Tr } (d\rho\, H).
\label{EWQ}
\end{equation}
In the step i) the state of the system evolves from  the complete mixture $\rho(t_0)= I/2$ to the pure state $|s\rangle\langle s|$. No work is performed 
on the system. As the energy levels remain degenerated there is no heat exchange but 
the decrease of entropy is compensated by the entropy increase in the 
measurement device and the environment. In the step ii) the state remains the same and again no work is performed. Similarly, no heat is exchanged and the entropy remains the same. During the step iii) the system equilibrates at any moment and the work
$W \leq kT \ln 2$ is adiabatically extracted from the bath and the entropy grows to its maximal value. The system ends the cycle again in the state $\rho(t_0)= I/2$. To avoid the conflict with the Second Law
we have to conclude that the following Landauer's principle for measurement (LPM) holds :\\
\emph{ A completion of a binary measurement, including reseting of a measuring device needs at least $kT\ln 2 $ of work.}\\
\par
{\bf Remarks} \emph{ It is often claimed that the work (at least $kT\ln 2$) needed to perform a binary measurement is actually used to erase a bit of  information in a "memory" of a measuring device \cite{Bennett}. Analogically, one can estimate by $\sim kT$  a minimal energy cost of any irreversible elementary gate. To support this picture  microscopic models of erasure have been proposed involving certain \emph{entropy-energy balance} \cite{Piech}. However, as shown in the Appendix II this argument is generally not convincing. Therefore, one is left with  phenomenological arguments as presented above which do not depend on the detailed model of the measurement procedure .} 
\par
Notice, that the arguments presented above do not apply if
both states $|\pm 1 \rangle$ are equilibrium ones in the sense of the definition of above, i.e. their relaxation times to the unique Gibbs state are exponentially long \footnote{A multitude of metastable states for a glassy system must be also treated as an equilibrium state, because the determination of an energy landscape is a computationally hard problem what makes impossible to design a cyclic process of work extraction from a heat bath.}. \\
\par 
Therefore, one faces the following alternative : \\
1) The LPM is  valid for all measurements, and therefore is not a consequence of the Second Law, but must be added as an independent additional postulate.\\
2) The LPM is a consequence of the Second Law and therefore does not need to hold for  equilibrium states. 
\par
The second possibility is much more likely as the laws of thermodynamics seem to provide ultimate, model independent limitations on physical processes. It seems that the task which does not violate the laws of thermodynamics can be realized in principle by a certain physical process. Therefore, one can formulate the plausible hypothesis which does not violate the laws of thermodynamics:\\
\textbf{Hypothesis I}\\
\emph{Equilibrium (metastable) states can be distinguished (measured), with the error decreasing exponentially with the size of the system, at the arbitrarily low energy cost.}\\
\par
Hypothesis I makes distinction between the information encoded in relaxing states or metastable states. Only for the former the associated information gain has a physical meaning and should be included into the thermodynamical entropy balance. This information is unstable and therefore must be recorded while the stable information need not. The crossover between them is described by the error behavior. For smaller systems the error grows and the "cost-free" acquiring of information must be replaced by the "costly" recording procedure.

\section{Perpetuum mobile based on quantum memory}

A scalable quantum memory for a single qubit is a system which consists of $N$ microscopic constituents (e.g. atoms, spins,..) interacting with a heat bath at the temperature $T>0$. The system is designed in such a way that the information bearing degrees of freedom form a quantum subsystem (encoded qubit) described by  2-dimensional Hilbert space spanned by the vectors of the form $|\psi\rangle \equiv|\psi\rangle\otimes |\omega_R\rangle$, with $|\omega_R\rangle$ being a purification of the fixed thermal equilibrium state for all other degrees of freedom. 
\par
Under the \emph{Hypothesis I}, the minimal needed assumptions concerning  operations on the single-qubit quantum memory are the following:\\
I) The eigenstates of $\sigma^z$ and $\sigma^x$ are equilibrium ones (i.e. metastable with life-times exponentially long in $N$).\\
II) One can perform effectively and "cost-free"  measurements of the observable $\sigma^z$,\\
III) The observables $\sigma^x ,\sigma^y$ and $\sigma^z$ can be implemented effectively to construct an interaction Hamiltonian with a relaxing qubit described by Pauli observables $X,Y,Z$ 
\begin{equation}
H_{\mathrm{int}}= \sigma^x \otimes X +\sigma^y \otimes Y +\sigma^z \otimes Z .
\label{coupling}
\end{equation} 
Here again "effectively" means that one needs time at most polynomial in $N$. 
\par
One can design now the following cyclic process which effectively extracts work from a heat bath using such quantum memory. The process consists of:\\
A) measuring the memory observable $\sigma^z$,\\
B) switching on the coupling Hamiltonian (\ref{coupling}),  performing a SWAP operation \cite{swap}, and  transferring the post-measurement memory state to the relaxing qubit,\\
C) extracting $kT\ln 2$ work from the bath using the knowledge of the state of the relaxing qubit and applying the procedure described in Section 3.
\par
In this process one uses the memory to reset the relaxing qubit into a known state without spending work.
Therefore, the net effect of this cyclic process is a subtraction of heat from the bath and the production of work what violates the Second Law of Thermodynamics.
\par
{\bf Remark} \emph{The example of above does not contradict the known results concerning the existence of metastable encoded qubit observables is some spin models like 4D-Kitaev model \cite{AlickiH3-Kitaev}. Namely, those observables are not given in terms of self-adjoint operators which can be effectively implemented and used as ingredients of the interaction Hamiltonian (\ref{coupling}), but are defined in term of measurement procedures accompanied by certain computational classical algorithms.}
\section{Stability v.s. reversibility}
The conclusion  from the \emph{Gedankenexperiment} discussed above raises a question which of the assumptions I, II, III concerning the properties of quantum memory are internally inconsistent. It seems that there exists a fundamental conflict between the stability of  states and possibility of applying reversible operations (gates) changing such states. Indeed, in the scheme discussed above one uses a reversible, unitary SWAP gate applied to stable states of the memory. Our experience based on the 4D-Kitaev models suggests that there is a common mechanism of state stabilization in the classical and quantum domains. It involves  free-energy barriers separating different metastable states which have to be overcame in the process of performing gate. Such an operation costs work which is then dissipated into environment. Therefore, the gates performed on stabilized states must be irreversible (non-Hamiltonians) transformations. This does not impair classical digital information processing where all practically used gates are irreversible but is a serious obstacle in the case of quantum information. Notice, that also classical reversible computation would be sensitive to chaos what implies that the classical analog computation could not overpower the digital one.
\par
One can make the statements of above more precise assuming that the work $W_g$ invested in the irreversible  gate is of the order of the free-energy barrier $F_N$ protecting information carrying states. Here, $N$ denotes the number of microscopic constituents of the information carrier (atoms, electrons, spins, etc.) and typically $F_N \sim N$. The same factor $F_N$ determines the stability of  protected states with respect to thermal noise characterized by their life-time 
\begin{equation}
\tau_N \simeq \tau_0 \exp\Bigl(\frac{F_N}{kT}\Bigr)\ \ , N >> 1
\label{lifetime}
\end{equation} 
where $\tau_0$ is a typical microscopic relaxation time scale. One can formulate the following \emph{Hypothesis II} which gives a more realistic estimation of the thermodynamical cost of irreversible information processing than the standard one based on the Landauer's Principle.\\
\textbf{Hypothesis II}\\
\emph{In order to perform a gate on the protected state one needs the amount of work}
\begin{equation}
W_g \simeq kT \ln\Bigl(\frac{\tau_N}{\tau_0}\Bigr)\simeq kT\, N .
\label{gate_cost}
\end{equation} 
\emph{which is dissipated into environment}.
\par
Actually, the zero-temperature analog of the formula (5) ($kT$ is 
replaced by  quantum fluctuations) is derived in the Appendix III for a 
spin-boson model and the presented derivation can be easily generalized 
to finite temperatures. Therefore, the Hypothesis II
is in fact a plausible conjecture that (5) is valid also for more 
sophisticated models.
\section{Conclusions}
The discussion of \emph{Gedankenexperiment} shows that a scalable quantum memory could be used as a perpetuum mobile of the second kind and hence cannot be realized in Nature. The fundamental assumptions behind the analysis of this model is that the properly formulated laws of thermodynamics are valid and the physical processes which are not forbidden by them can be realized. Those general principles suggest the alternative hypothesis concerning the thermodynamical cost of acquiring information and performing operations for the case of stabilized information carriers. The conflict between stability of information and reversibility of gates does not restrict the irreversible classical digital information processing but suggests unfeasibility of large scale quantum information one. The heuristic arguments presented in this paper can be generally accepted only if there are supported by a large body of independent theoretical and experimental evidence. Therefore, the analysis of microscopic models of the candidates for quantum memory  is still important and will be for  sure continued in the near future. 

\textbf{Acknowledgments}
The author thanks Hector Bombin, Micha\l\ Horodecki, Charles Bennett, Daniel Lidar and Stanis\l aw Kryszewski for discussions. The support by the Polish Ministry of Science and Higher Education, grant NN 202208238 is acknowledged.  
\par
\textbf{Appendix I. Markovian model reproducing the laws of thermodynamics}
\par
The laws of thermodynamics can be derived from the following model of the open system 
coupled to several heat baths and controlled by external forces \cite{AL1979}. The density matrix of the system $\rho(t)$ satisfies the following Markovian Master Equation
\begin{equation}
\frac{d}{dt}\rho(t) = -i[H(t),\rho(t)] + \sum_j \mathcal{L}_j(t)\rho(t),
\label{MME}
\end{equation} 
where for any $ 0\leq t\leq\infty$ $\mathcal{L}_j(t)$ is a generator of a completely positive dynamical semigroup which satisfies
\begin{equation}
\mathcal{L}_j(t)\rho_j^{eq}(t)=0\  ,\ \rho_j^{eq}(t) = \frac{e^{-\beta_j H(t)}}{\mathrm{Tr}e^{-\beta_j H(t)}}\ .
\label{eq}
\end{equation} 
Here $\beta_j= 1/kT_j$ is the inverse temperature of the $j$-th heat bath. The equation of motion (\ref{MME})(\ref{eq}) can be derived from a microscopic Hamiltonian model using the weak coupling assumption and for slowly varying external fields \cite{DavSp}. \\
The First Law of thermodynamics becomes now the definition of work $W$ performed on a system and heat $Q$ absorbed by the system with the obvious definition of the internal energy $E$
\begin{equation}
E(t) = \mathrm{Tr}\bigl(\rho(t)H(t)\bigr)\ ,\ 
\frac{d}{dt}W(t) = \mathrm{Tr}\bigl(\rho(t)\frac{dH(t)}{dt}\bigr)\ ,
\label{work}
\end{equation} 
\begin{equation}
 \frac{d}{dt}Q(t) = \mathrm{Tr}\bigl(\frac{d\rho(t)}{dt}H(t)\bigr)= \sum_j\mathrm{Tr}\bigl(H(t)\mathcal{L}_j(t)\rho(t)\bigr)\equiv \sum_j\frac{d}{dt}Q_j(t)\ .
\label{heat}
\end{equation} 
where $Q_j$ is the heat absorbed by the system from the $j$-th bath.\\
Defining the entropy as $S(t)= -k \mathrm{Tr}\bigl(\rho(t)\ln \rho(t)\bigr)$ one obtains the Second Law 
\begin{equation}
\frac{d}{dt}S(t) - \sum_j\frac{1}{T_j} \frac{d}{dt}Q_j(t)= \sum_j\sigma_j(t)\geq 0 
\label{IIlaw}
\end{equation} 
where the entropy production caused by the $j$-baths is given by 
\begin{equation}
\sigma_j(t)= k \mathrm{Tr}\bigl(\mathcal{L}_j(t)\rho(t)[\ln \rho(t) - \ln \rho_j^{eq}(t)]\bigr)\geq 0
\label{enprod}
\end{equation} 
and its positivity follows from (\ref{eq}) and complete positivity of the map 
$\exp\{s\mathcal{L}_j(t)\}$, $s\geq 0$.
\par
\textbf{Appendix II. Argument based on energy and entropy balance}
\par
Quite often the microscopic derivations of Landauer's principle are based on the following picture.
We consider a process with an initial and final product states for a system coupled to a heat bath
\begin{equation}
\rho_{in}\otimes\omega (\beta)\rightarrow \rho_{fin}\otimes\omega' 
\label{inst1}
\end{equation} 
where $\omega (\beta)$ is a Gibbs state of a bath at the inverse temperature $\beta$ and $\omega'$ is a final state of a bath, not necessarily given by another Gibbs state.\\
Using the definitions and notation
\begin{equation}
\omega (\beta)= \frac{e^{-\beta H_{bath}}}{Z(\beta)}\ , E(\beta)= \mathrm{Tr}\bigl(\omega (\beta)H_{bath}\bigr)\ ,\ S(\beta)=-k\mathrm{Tr}\bigl(\omega (\beta)\ln \omega (\beta)\bigr) 
\label{inst}
\end{equation} 
one can easily compute
\begin{equation}
\frac{d}{d\beta} S(\beta) = k\beta\frac{d}{d\beta} E(\beta)\ .
\label{enbal}
\end{equation} 
From (\ref{inst1}) and the fact that the total system is an isolated Hamiltonian one, the entropy balance follows 
\begin{equation}
S(\rho_{fin})- S(\rho_{in})= S(\beta) - S(\beta')\simeq k\beta \bigl(E(\beta) - E(\beta')\bigr)
\label{enbal2}
\end{equation} 
where $\beta'$ is the inverse temperature of the Gibbs state such that $S(\beta')= -k\mathrm{Tr}(\omega'\ln\omega')$.
In the last equality in (\ref{enbal2})one uses the fact that a heat bath is a large physical system and its interaction with a small open system can only infinitesimally change bath's intensive parameters what implies
$\beta'\simeq\beta$. As the Gibbs state minimizes internal energy under the condition of a fixed entropy the energy gain of a bath satisfies
\begin{equation}
\Delta E = \mathrm{Tr}(\omega' H_{bath}) - E(\beta)\geq E(\beta') - E(\beta)\ .
\label{enbal3}
\end{equation} 
Defining 
\begin{equation}
\Delta S = S(\rho_{fin})- S(\rho_{in})= S(\beta) - S(\beta')
\label{dent}
\end{equation} 
and using (\ref{enbal2}),(\ref{enbal3}) one obtains the inequality
\begin{equation}
\Delta E + T \Delta S \geq 0\ .
\label{enen}
\end{equation} 
One can apply now the scheme of above to a model of reseting a single bit of information in a memory of measuring device.
The bit is supported by two degenerated eigenstates of the memory $|0\rangle, |1\rangle$ . The initial state encodes an unknown bit what corresponds to the initial state $\rho_{in} = 1/2(|0\rangle\langle 0| + |1\rangle\langle 1|)$ with the entropy $k\ln 2$, and the final state is a fixed reference state, say $|0\rangle$, with the entropy equal to $0$. Therefore, using (\ref{enen}) one obtains the lower bound for the increase of the bath's internal energy
\begin{equation}
\Delta E \geq kT\ln 2\ .
\label{lbound}
\end{equation} 
As the process is cyclic in the sense that the external time-dependent control fields switched on at the beginning of the process are switched off  at its end, and the energy of the $2$-level system is not changed one can attribute $\Delta E$ to the amount of work performed by the external forces and dissipated into the bath's degrees of freedom. Hence, the Landauer's principle for bit's erasure seems to be justified on the microscopic basis. Moreover, as an actual measurement which transforms the initial reference state $|0\rangle$ into $|0\rangle$ or $|1\rangle$ does not change the entropy this part of a cyclic measurement process does not need work and hence the LPM seems to be valid as well.
\par
Unfortunately, the arguments of above are not convincing. The main problem is the entropy balance based on the assumption of the exact product structure for initial and final states (\ref{inst1}). Indeed, a weak coupling to a heath bath suggests an \emph{approximative} product state structure but due to the \emph{discontinuity of entropy} in the limit of large systems 
we cannot use it for the estimation of entropy. This fact is 
expressed in terms of Fannes inequality \cite{Fan} for two close density matrices of a system with $D$-dimensional Hilbert space and $\|\cdot\|_1$ denoting the trace norm
\begin{equation}
|S(\rho)- S(\rho')| \leq \|\rho -\rho'\|_1 \ln D - \|\rho -\rho'\|_1\ln(\|\rho -\rho'\|_1)\ .
\label{fan}
\end{equation}
To illustrate this problem one can consider the model discussed in the Appendix I. The validity of the Markovian approximation means that the state of the total system is well-approximated by the product
$\rho(t)\otimes \omega_{B}$ where $\rho(t)$ is a solution of the Master equation (\ref{MME}), and $\omega_B$ is a fixed stationary state of the bath. Obviously, this product form is not consistent with the constant entropy of the total Hamiltonian system. The missing entropy is hidden in the small correction terms describing the residual system - bath correlations  and local perturbations of the bath's state which practically do not influence the values of measured observables.  Therefore, to obtain a proper entropy, heat and work balance one has to use their definitions as presented in Appendix I and needs an equation of motion for a system like that obtained in the Markovian limit (\ref{MME}).
\par
\textbf{Appendix III. A quantum model of stable information carrier}
\par
The model represents a system with two degenerate ground states which are "macroscopically distinguishable" and stable and hence can be used as a single bit memory, in particular, as an element of a quantum measurement apparatus which records the value of a dichotomic observable. The system consists of a spin-1/2 described by the Pauli matrices coupled to a "macroscopic" system  defined in terms of bosonic fields $a(\omega), a^+(\omega)$
satisfying canonical commutation relations
\begin{equation}
[a(\omega), a^+(\omega')]= \delta (\omega - \omega')\ ,\ \omega,\omega' 
\in [0 ,\infty)
\label{1}
\end{equation}
The system Hamiltonian is a simple version of the spin-boson Hamiltonian studied, for example, in \cite{Alicki:2006}
\begin{equation}
H = H_0 +\sigma^z\otimes \int_0^{\infty}d\omega \,\omega\bigl( {\bar g}(\omega) a(\omega)
+ g(\omega) a^+(\omega)\bigr)
\label{2}
\end{equation}
with
\begin{equation}
H_0 = \int_0^{\infty} d\omega\,\omega\,a^+(\omega)a(\omega) .
\label{3}
\end{equation}
The unitary "dressing" operator defined by
\begin{equation}
U_{d} = \exp\Bigl\{\sigma^z\otimes \int_0^{\infty} d\omega ({\bar g}(\omega) a(\omega)
- g(\omega) a^+(\omega))\Bigr\}
\label{3a}
\end{equation}
diagonalizes the system Hamiltonian, i.e. $U_{d}H U_{d}^{\dagger} = H_0$ and therefore makes the model exactly solvable.
In particular one can find two degenerate ground states of $H$
\begin{equation}
H |\psi_{\pm}\rangle = -E_g |\psi_{\pm}\rangle 
\label{4}
\end{equation}
where 
\begin{equation}
|\psi_{\pm}\rangle = |\pm\rangle\otimes|[\pm g]\rangle \ ,\ \sigma^z |\pm\rangle = \pm |\pm\rangle
\label{5}
\end{equation}
and $|[f]\rangle$ denotes the field coherent state obtained from the vacuum state $|\Omega\rangle$ by the action of the 
Weyl operator $W[f]$, i.e.
\begin{equation}
|[f]\rangle = W[f]|\Omega\rangle,\   W[f]=\exp\Bigl\{\int_0^{\infty} d\omega ({\bar f}(\omega) a(\omega)- f(\omega) a^+(\omega)\Bigr\}.
\label{6}
\end{equation}
The "classicality" or "macroscopicality" of the field states  $|[\pm g]\rangle$ is characterized by the averaged number of bosons $N$ and the energy of ground states $-E_g$ 

$$
N= \int_0^{\infty} |g(\omega)|^2 d\omega ,\ E_g = \int_0^{\infty} \omega |g(\omega)|^2 d\omega\ .
\eqno(7)
$$
For $N >>1$ the coherent states of the field corresponding to "classical field configurations"  $\pm g$ are those macroscopic "pointer states" which allow directly, without any cost, to determine  which of the two ground states is occupied by the system. The error  of this distinguishing process is exponentially small in $N$ and given by the overlap of coherent states
\begin{equation}
\epsilon = |\langle [g]|[-g]\rangle|^2 = e^{-4N}.
\label{8}
\end{equation}
The spin-boson system can be used as a carrier of a bit or an element of the quantum measurement device which records the measurement result. In both cases one applies a fast NOT (or CNOT) gate on the spin part, which produces a new spin-boson state from one of the initial ground states, say $|\psi_{+}\rangle$
\begin{equation}
|{\psi\prime}_{+}\rangle = \sigma^x|{\psi}_{+}\rangle =|-\rangle\otimes|[g]\rangle 
\label{9}
\end{equation}
with the averaged energy $\langle{\psi\prime}_{+}|H|{\psi\prime}_{+}\rangle = 3E_g$. The difference between this energy and the ground state one (energy barrier) is equal to the work $W_g = 4E_g$  needed to implement a NOT gate by a suitable time-dependent Hamiltonian. One can exactly compute the subsequent evolution of the spin-boson state
\begin{equation}
|{\psi\prime}_{+}(t)\rangle = e^{-iHt}|{\psi\prime}_{+}\rangle = e^{i\alpha(t)}|-\rangle\otimes W[2g_t - g]|\Omega\rangle  
\label{10}
\end{equation}
where $\alpha(t)$ is an irrelevant phase and $g_t(\omega) = e^{-i\omega t}g(\omega)$ is a traveling wave. For long $t\to\infty$ the classical traveling wave $2g_t$ becomes orthogonal to the initial bounded field $-g$ and the state possesses a product structure \footnote{This tensor product structure corresponds to the property of Fock spaces 
$\mathcal{F}({\mathcal{H}}'\oplus{\mathcal{H}}'')=\mathcal{F}({\mathcal{H}}')\otimes\mathcal{F}({\mathcal{H}}'')$ 
with the corresponding identification of vacuum states $|0\rangle \equiv 
|0'\rangle\otimes|0''\rangle$.}
\begin{equation}
|{\psi\prime}_{+}(t)\rangle \sim |{\psi}_{-}\rangle\otimes |\psi[2g_t ]\rangle . 
\label{11}
\end{equation}
The component $|\psi[2g_t ]\rangle$ carries the energy $4E_g$ to infinity describing its dissipation into environment.
The total state $|{\psi\prime}_{+}(t)\rangle $ becomes indistinguishable from the ground state $|{\psi}_{-}\rangle$ by any local measurement. Hence, one obtains an irreversible  NOT gate performed on the stable states encoding a bit of information and the thermodynamical cost of the gate is $4E_g$ of work. Notice, that when the system is applied to record a measured state one does not need to reset it to a fixed reference state. Namely, the observed \emph{change} of the spin-boson ground state could be an indicator of a one of two states of the measured system.
\par
It is instructive to see how the dissipation mechanism of above prevents the encoding of a qubit state in our memory device. Consider a fast SWAP gate between a certain qubit at the initial state $\alpha|0\rangle + \beta|1\rangle$ and the spin degree of freedom of our memory at the initial  total memory state $|{\psi}_{+}\rangle = |+\rangle\otimes|[g]\rangle$. The state of the total system, just after SWAP gate, is given by
\begin{equation}
\bigl(\alpha|-\rangle +\beta|+\rangle\bigr)\otimes |[ g]|\rangle \otimes |0\rangle .
\label{12}
\end{equation}
For long enough times the total state evolves into
\begin{equation}
\Bigl[\alpha|-\rangle\otimes |[-g]\rangle\otimes|[2g_t]\rangle  +\beta|+\rangle\otimes |[g]\rangle\otimes|\Omega\prime\rangle\Bigr]\otimes|0\rangle
\label{13}
\end{equation}
where one uses the fact that $\lim_{t\to\infty}\langle g|g_t\rangle = 0$ and $|\Omega\prime\rangle$ denotes the vacuum state of the field degrees of freedom which  represent the traveling waves with support far away from the localized states of the memory. Those traveling waves form the environment of the memory which can be traced out to give the following asymptotic state of the memory itself
\begin{equation}
\rho_M = |\alpha|^2 |\psi_{-}\rangle\langle\psi_{-}| + |\beta|^2 |\psi_{+}\rangle\langle\psi_{+}| + \Bigl[\alpha\bar{\beta}\langle\Omega\prime|[2g_t]\rangle|\psi_{-}\rangle\langle\psi_{+}| + \mathrm{h.c}\Bigr].
\label{14}
\end{equation}
As the off-diagonal terms are exponentially small ($|\langle\Omega\prime|[2g_t]\rangle|= e^{-2N}$) the state (\ref{14}) is a mixture of two stable memory states. It means that the decoherence process accompanying the dissipation of $4E_g$ of energy destroys quantum coherence of the swapped qubit state.\\
The similar analysis of the measurement/encoding process in the case of finite temperature will be presented in the forthcoming paper.

\end{document}